\documentclass[12pt]{article}

\usepackage{epsfig}

\usepackage{amssymb}
\usepackage{amsmath}
\usepackage{amsfonts}

\def\be{\begin{equation}}
\def\ee{\end{equation}}
\def\ba{\begin{array}{c}}
\def\ea{\end{array}}

\def\ben{$$}
\def\een{$$}

\newcommand{\bea}{\begin{eqnarray}}
\newcommand{\eea}{\end{eqnarray}}

\newcommand{\kt}{\rangle}

\begin{document}

\titlepage

\vspace{.35cm}

 \begin{center}{\Large \bf

Quantum Big Bang without fine-tuning in a toy-model

  }\end{center}

\vspace{10mm}

 \begin{center}

 {\bf Miloslav Znojil}

 \vspace{3mm}
Nuclear Physics Institute ASCR,

250 68 \v{R}e\v{z}, Czech Republic

{e-mail: znojil@ujf.cas.cz}

\vspace{3mm}

\end{center}


\vspace{15mm}

\section*{Abstract}

The question of possible physics before Big Bang (or after Big
Crunch) is addressed via a schematic non-covariant simulation of the
loss of observability of the Universe. Our model is drastically
simplified by the reduction of its degrees of freedom to the mere
finite number. The Hilbert space of states is then allowed
time-dependent and singular at the critical time $t=t_c$. This
option circumvents several traditional theoretical difficulties in a
way illustrated via solvable examples. In particular, the unitary
evolution of our toy-model quantum Universe is shown interruptible,
without any fine-tuning, at the instant of its bang or collapse $t=
t_c$.

\newpage

\section{Introduction}

The -- apparently purely philosophical -- question of ``what did
exist before the Big Bang?" has recently changed its status. Its
numerous recent innovative and non-speculative treatements may be
sampled, e.g., by the Penrose's deep theoretical analysis of
possible physics before Big Bang \cite{Penrose} or by the
Gurzadyan's and Penrose's proposal of the existence of cyclically
recurring ``aeons" before Big Bang, with potentially measurable
(i.e., in principle, falsifiable!) consequences. Naturally, the
topic involves also the parallel question of possible scenarios of
the evolution of the Universe after the Big Crunch, i.e., at $t>
t_{final}$ \cite{PenroseII}.

One of the main difficulties encountered in similar considerations
can be seen in the fact that our current knowledge of the laws of
nature is not too well adapted to the description of the Universe
near the Big Bang (i.e., schematically, in a short interval of times
$t \approx t_c=t_{initial}$) or, if you wish, near the Big Crunch
(i.e., at $t \approx t_c=t_{final}$). At the same time, the picture
offered by the {\em classical} theory of general relativity seems
compatible with the schematic, simplified but still intuitively
acceptable scenario in which the existence of the critical
Big-Bang/Big-Crunch (BBC) instant $t=t_c$ may be visualized as the
time-dependence of {\em any} $N-$plet of the spatial grid-point
coordinates $g_j(t)$, $j=1,2,\ldots,N$ (or of their, in principle,
measured distances in a suitable frame) with the complete-confluence
property
 \be
 \lim_{t \to t_c}g_{j}(t)=g_c
 \,, \ \ \ \ \ \ j = 1,2,\ldots,N\,.
 \label{lide}
 \ee
The key difficulties emerge when one tries to make this picture
compatible with the requirements of quantum theory. In this context,
Penrose \cite{Penrose} emphasized that whenever one tries to
``quantize" the picture treating the grid points $g_j(t)$ (or any
other measurable data) as eigenvalues of an {\it ad hoc}
self-adjoint operator ${\cal O}={\cal O}^\dagger$ in Hilbert space
${\cal H}$, one encounters the well-known fine-tuning problem.
Indeed, near $t=t_c$ it becomes {\em extremely difficult} to
suppress, by the fine-tuning of parameters, the {\em generic} and
well known property of the eigenvalues of {\em any} self-adjoint
${\cal O}={\cal O}^\dagger$ which tend to avoid their crossings near
{\em any} point of potential degeneracy.

The recent proposal of a  conformal cyclic cosmology \cite{CCC} may
be perceived as one of the possible ways out of this
quantum-theoretical trap. One simply admits that the $t=t_c$
degeneracy (\ref{lide}) remains avoided and that the
avoided-crossing nature of the Big Bang {\em must} leave its traces,
e.g., in the emergence of certain concentric circles in the cosmic
microwave background measured by the Wilkinson Microwave Background
Probe.

In our present paper we intend to join the discussion by showing
that even in the framework of the entirely standard quantum theory
the alternative assumption of the {\em unavoided} degeneracy of
eigenvalues at the Big Bang [as required, say, by Eq.~(\ref{lide})]
{\em need not necessarily} require any low-probability fine-tuning.

The conceptual core of such a message may be traced back to our
recently proposed extension of the quantum-theoretical perspective
(cf. paper \cite{timedep} or more detailed exposition \cite{SIGMA})
which does not modify any ``first principles" of quantum theory. One
merely decides to work with the manifestly time-dependent
representation of the ``standard" physical Hilbert space of states,
${\cal H}= {\cal H}^{(S)}(t)$, which may simply cease to exist at
$t=t_c$.

The latter option is to be shown here to enlarge the number of free
parameters in the corresponding quantum models of dynamics in such a
manner that one can satisfy the degeneracy constraints of the form
(\ref{lide}) without any true difficulties. In addition, an optimal
balance may be also achieved between the``classical" and ``quantum"
input information about the dynamics of the model.

For the sake of simplicity of presentation of the idea just an
elementary illustrative phenomenological quantum model of
Sec.~\ref{themodel} will be considered. In particular, no time
re-parametrization invariance will be implemented to lead to an
analog of the Weeler-DeWitt equation. In this way, in particular,
the initial/final time moments  will stay finite rather than
transferred into conformal infinities.

The detailed analysis or our model will enable us to demonstrate
that the BBC-like degeneracies of eigenvalues {\em need not
necessarily} induce any enhanced sensitivity to perturbations nor
the need of any particular fine-tuning. Thus, in our schematic model
the quantum Universe may become strictly unobservable both before
$t= t_{initial}$ and after $t= t_{final}$.

The technical essence of our message will lie in the recommended use
of adiabatically time-dependent inner products in the Hilbert space
of quantum theory (cf. Sec.~\ref{themethod}). In the main body of
the paper our quantum description of the BBC phenomenon will be
illustrated via several non-numerical, exactly solvable examples
(cf. Secs.~\ref{themeat} and \ref{thebeat} and Appendix A). In the
subsequent discussion in Sec.~\ref{therest} we shall emphasize that
in the close vicinity of the critical BBC times $t=
t_{initial/final}=t_c$ the role of the (adiabatic) time-dependence
of the Hilbert space proves crucial.

\section{The  model \label{themodel} }

For methodical purposes several drastic mathematical simplifications
of the overall physical scenario will be accepted. Firstly, we shall
start building the quantum states of our schematic Universe  inside
Hilbert space ${\cal H}^{(friendly)}$ of a finite dimension $N <
\infty$. Secondly, we shall consider quantum theory of pure states
only (i.e., no statistical physics). Thirdly, we shall follow some
preliminary considerations by B\'{\i}la \cite{Bila} and treat the
time-evolution of wave functions $|\psi(t)\kt$ as adiabatic,
circumventing thereby several technical complications as listed and
discussed in \cite{timedep}.

Last but not least, we shall accept here a very pragmatic attitude
towards the (up to now, unresolved) theoretical conflict between
quantum theory and general relativity. In this conflict we shall
never leave the standard textbook quantum mechanics in its
cryptohermitian or three-Hilbert-space (THS) recent reformulations
as summarized, e.g., in our compact review \cite{SIGMA}. We believe
that for the time scales chosen as extremely short, this constraint
(leading, e.g., to the manifest violation of the covariance
requirements) may still represent a more or less safe territory of
valid and consistent theoretical considerations admitting subsequent
amendments, in principle at least.

For our present purposes the quantized generator of the time
evolution (i.e., our toy-model Hamiltonian operator $H$) will be
chosen in the following, extremely schematic and purely kinetic real
and symmetric $N$-by-$N$-matrix time-independent and force-free form
 \be
 H =H^{(N)}=
  \left[ \begin {array}{cccccc}
   2&-1&0&\ldots&0&0
\\-1&2&-1&\ddots&\vdots&\vdots
 \\
 0&-1&2&\ddots&0&0
 \\
 0&0
&\ddots&\ddots&-1&0
 \\{}\vdots&\vdots&\ddots&-1&2&-1
 \\{}0&0&\ldots&0&-1&2\\
 \end {array} \right] \,.
 \label{kinetie}
 \ee
For the questions we are going to ask (and concerning, e.g., the
observability nature of the ``eligible histories" of our schematic
``Universe" near its BBC singularities) this operator itself even
cannot be interpreted as {\em directly} related to the existence of
these singularities. The reason is that precisely the very dynamical
source of the emergence of these singularities lies already {\em
beyond} the above-selected quantum-mechanical short-times scope and
methodical range of our present message.

In the resulting picture of reality near the critical time $t=t_c$
all the information about the physics of the BBC dynamics will be
assumed {\em given in advance} (say,, from the purely external
sources offered by the cosmological model-building and/or by
non-quantum general relativity). We shall only work here with an
empty-space {\em phenomenological} model of the collapsing Universe
near  $t=t_c$.

The spatial or geometric structure of the collapse will lie in the
center of our interest. Four our purposes it will be characterized
by the measurability and/or measurements of a finite sample
$g_1(t)$, $g_2(t)$, \ldots, $g_N(t)$ of the $N$ spatial grid points
{\em at a classical, continuous time}. These representative
grid-point real coordinates will be treated as eigenvalues of a
certain pre-determined general-matrix operator of the  most
essential observable
 \be
 \hat{G} =\hat{G}^{(N)}(t)=
  \left[ \begin {array}{cccc}
   \gamma_{11}&\gamma_{12}&\ldots&\gamma_{1N}
 \\
   \gamma_{21}&\gamma_{22}&\ldots&\gamma_{2N}
 \\
   \ldots&\ldots&\ldots&\ldots
 \\
   \gamma_{N1}&\gamma_{N2}&\ldots&\gamma_{NN}
 \end {array} \right] \,.
 \label{potentialie}
 \ee
After Big Bang and before Big Crunch, the natural requirement of
observability of the Universe forces us to impose $N$ conditions of
reality of the spectrum of this operator (i.e., in our toy model, of
this matrix),
 \be
 {\rm Im}\ g_{j}(t)=0\,, \ \ \ \ j = 1,2,\ldots,N\,,
 \ \ \ \ t_{initial}\leq t \leq t_{final}\,.
 \label{acoll}
 \ee
Optionally, we might also add another, complementary requirement
guaranteeing {\em either} the partial {\em or} the complete
non-measurability of the space before Big Bang or after Big Crunch,
 \be
 {\rm Im}\ g_{j}(t)\neq 0\,, \ \ \ \ j = 1,2,\ldots,N_{BBC}\,,
 \ \ \ \ t \notin [t_{initial},t_{final}]\,,\ \ \ \ \
 N_{BBC}\leq N\,.
 \label{becoll}
 \ee
In this language the BBC phenomenon itself will be simulated just by
the $N-1$ conditions of a complete confluence of the $N-$plet of
eigenvalues
 \be
 \lim_{t \to t_c}g_{j}(t)=g_{N}(t_c)\,, \ \ \ \ j = 1,2,\ldots,N-1\,
 \label{collide}
 \ee
which would guarantee also the complete single-point geometrical
collapse of our toy-model Universe at the critical time.

Let us re-emphasize that we shall solely speak here about the
privileged (viz., time-evolution) boosts generated by the quantum
mechanical Hamiltonian operators $H$ and considered just along
certain very short intervals of the time which will be assumed
measured by the classical clocks. Naturally, such a decision
(motivated, first of all, by the technical feasibility of at least
some quantitative considerations) will force us to leave many
important (and, up to these days, open) questions entirely aside.

Due to these assumptions we shall be able to keep working with the
naive, non-covariant Schr\"{o}dinger time-evolution equation.
Naturally, we shall be unable to estimate the extent of the
modifications of this picture after some future (and, of course,
theoretically necessary) transition to the less scale-restrictive
scenarios based on some suitable general-relativistic covariance
requirements (sampled, e.g., by their well known
incorporation~\cite{Wuitt} by Bryce DeWitt).

Our present key message will be restricted, therefore, to the
constructive demonstration that in a very close vicinity of the BBC
regime the language of quantum mechanics admits the complete (or, in
alternative models, partial) loss of the measurability of the
geometry of the (collapsing) space before the Big Bang and/or after
the Big Crunch. In this sense, we do not see any {\em theoretical
necessity} of the existence of any measurable Universe (or,
alternatively, of a measurable Universe with the same number of
dimensions), say, before the Big Bang.

This being said, we should add, as early as possible, that our
present model is really too schematic for any cosmology-related
and/or prediction-making  purposes. In particular, the
quantum-mechanics-based demonstration of the {\em possibility} of
the (partial or complete) complexification of the coordinates (say,
after the Big Crunch) certainly does not exclude their subsequent
return to reality (say, in the cyclic form proposed in the very
interesting recent preprint~\cite{CCC}).

\section{The method \label{themethod}}

In a way inspired by the so called ${\cal PT}-$symmetric quantum
mechanics \cite{Carl} the key to the resolution of the
above-mentioned Penrose's paradox of incompatibility of the
assumption of Hermiticity of observables with the existence of the
critical BBC times $t_c$ will be sought here in the omission of the
former, overrestrictive assumption. In other words, we shall broaden
the class of the admissible operators of geometry
(\ref{potentialie}) and admit that
 \be
 \hat{G}(t)\neq
 \hat{G}^\dagger(t) \ \ \  {\rm in} \ \ \ {\cal H}^{(friendly)}\,.
 \label{assu}
 \ee
One must emphasize here that this relation {\em must not} be read as
a non-Hermiticity of $\hat{G}(t)$. Its true meaning  is much
simpler: Equation (\ref{assu}) will be understood as a mere
consequence of our re-classification of the original
time-independent representation ${\cal H}^{(friendly)}\neq {\cal
H}^{(friendly)}(t)$ of the Hilbert space of states as
overrestrictive and {\em manifestly unphysical}.

The necessary mathematics underlying such a change of perspective
has been offered in Refs.~\cite{Geyer}. The main idea is that the
naive choice of Hilbert space ${\cal H}^{(friendly)}$ is being
replaced by a more flexible option. In it, the inner product is
being determined via operator $\Theta=\Theta^\dagger>0$ called
metric (i.e., ``Hilbert-space" metric, certainly different from the
much more common Riemann-space-metric {\em function} $g_{\mu \nu}$).

Naturally, such a decision leads to the new form of Hermitian
conjugation (marked, conveniently, by a double-cross superscript
$^\ddagger$) and, hence, to the new, unitarily inequivalent Hilbert
space ${\cal H}^{(true)}$ which is {\em declared} physical. Any
pre-selected non-Hermitian operator acting in ${\cal
H}^{(friendly)}$ and possessing real spectrum may be then
reinterpreted as the ``cryptohermitian" \cite{Smilga} operator of an
observable quantity, i.e., as an operator which becomes self-adjoint
in the amended, physical Hilbert space ${\cal H}^{(true)}$.

A realistic BBC phenomenology may be built on this background. In
the simplest arrangement of the theory the transition from trivial
metric $\Theta=I:=\Theta^{(Dirac)}$ to nontrivial metric
$\Theta=\Theta(t)>0$ will in fact represent, in our present
considerations, the only difference between ${\cal H}^{(friendly)}$
and ${\cal H}^{(true)}$. Nevertheless, it is necessary to emphasize
that in contrast to the usual applications of transition from ${\cal
H}^{(friendly)}$  to ${\cal H}^{(true)}$ dealing with single
observable (usually, with the Hamiltonian), our present model will
require the {\em simultaneous} guarantee of cryptohermiticity of
{\em both} our observables  $H$ and $\hat{G}$.

In the light of property (\ref{assu}) of the latter operator one
really cannot choose $\Theta=\Theta^{(Dirac)}$ so that the
Hermiticity $H=H^\dagger$ of our toy Hamiltonian in ${\cal
H}^{(friendly)}$ is in fact irrelevant. The metric must be
constructed which would make {\em both} our operators of observables
self-adjoint, yielding
 \be
 H=
 H^\ddagger
 :=\Theta^{-1}\,
 H^\dagger\,\Theta \ \equiv\
 \Theta^{-1}\,
 H\,\Theta
 \,,
 \label{sue}
 \ee
as well as
 \be
 \hat{G}(t)=
 \hat{G}^\ddagger(t)
 :=\Theta^{-1}\,
 \hat{G}^\dagger(t)\,\Theta
 \,.
 \label{due}
 \ee
From the point of view of physics the additional model-building
freedom offered by Eqs.~(\ref{assu}), (\ref{sue}) and (\ref{due})
opens a way towards the construction of metrics which could vary
with time, $\Theta=\Theta(t)$. In this manner many ``no-go"
consequences of the restrictive formal framework provided by the
ill-chosen space ${\cal H}^{(friendly)}$ may be circumvented
\cite{timedep}.

The former constraint (\ref{sue}) appears much easier to satisfy
because our Hamiltonian itself remains time-independent. As long as
this operator is represented by the real and symmetric
$N-$dimensional matrix (\ref{kinetie}), the most natural
representation of the metric can be provided by polynomial formula
 \be
 \Theta(t)=a(t)\,I + b(t)\,H + c(t)\,H^2+ \ldots + z(t)\,H^{N-1}\,
 \label{anza}
 \ee
containing $N$ unknown real-function coefficients. Such an ansatz
may be inserted in Eq.~(\ref{due}) yielding the ultimate set of
algebraic constraints expressed in terms of modified commutators
$[A,B]_\dagger := AB-B^\dagger A$,
 \be
 a(t)\,[I,\hat{G}(t)]_\dagger +
 b(t)\,[H,\hat{G}(t)]_\dagger + c(t)\,[H^2,\hat{G}(t)]_\dagger+ \ldots +
 z(t)\,[H^{N-1},\hat{G}(t)]_\dagger= 0\,.
 \label{ultimo}
 \ee
We may summarize that the variability of the adiabatically
time-dependent real matrices (\ref{potentialie}) carrying the input
dynamical information and containing $N^2$ independent matrix
elements $\gamma_{ij}(t)$ is only restricted by the $N(N-1)/2$
metric-compatibility conditions (\ref{ultimo}), by the $N$
spectral-reality (i.e., Universe-observability) conditions
(\ref{acoll}) and by the $N-1$ complete-degeneracy conditions
(\ref{collide}) imposed at $t=t_c$. This means that at least the
$(N-1)(N-2)/2-$plet of input parameters remains arbitrary. No
particular fine-tuning will be needed at $N \geq 3$, therefore.

Naturally, the domain of variability of the input parameters is not
arbitrary since one must guarantee the invertibility and positive
definiteness of the metric as well as its compatibility with the
concrete pre-selected Hamiltonian $H$. Via Eq.~(\ref{ultimo}) these
conditions further restrict the variability of parameters in
$\hat{G}(t)$ and in $\Theta(t)$ to a certain domain ${\cal
D}^{(physical)}(t)$. Although the exhaustive specification of this
time-dependent domain is difficult in general, it is usually
sufficient and not so difficult to find its nonempty
time-independent subdomain ${\cal D}^{(practical)}$. Better insight
in the latter restrictions may be gained via the detailed inspection
of the model at the lowest dimensions~$N$.

\section{Metrics $\Theta^{(N)}$ \label{themeat} }

\subsection{Grid dimension $N=2$}

The Hamiltonian as well as the metric are elementary at $N=2$,
 \be
H^{(2)} =
 \left (
 \begin{array}{cc}
 2&-1\\
 -1&2
 \ea\,
 \right )\,,\ \ \ \
 \Theta^{(2)}(t)=
 \left (
 \begin{array}{cc}
 a(t)+2b(t)&-b(t)\\
 -b(t)&a(t)+2b(t)
 \ea\,
 \right )\,.
 \label{onidva}
 \ee
The eigenvalues $\theta_{\pm}(t)=a(t)+2b(t)\pm b(t)$ of the metric
are easily evaluated. The positivity of the metric (i.e., of all of
its eigenvalues) imposes just the single constraint at $N=2$, viz.,
$ a(t)>\max(-b(t),-3b(t))$. Inside this interval the standard
probabilistic interpretation of our $N=2$ quantum Universe is
guaranteed.

The detailed dynamics of the model must be deduced (typically, via
the principle of correspondence) from the classical theory of
gravity. In our approach this information is carried solely by the
operator of geometry (\ref{potentialie}). Even its $N=2$ realization
illustrates quite well the idea. We even do not need the fully
general matrix for this purpose. One of its elements may certainly
be fixed by the convenient location of the BBC limiting coordinate
in the origin, $g_c(t_c)=0$. The resulting reduced three-parametric
matrix
 \be
 \hat{G}^{(2)}(t)=
 \left (
 \begin{array}{cc}
 -r(t)&-v(t)\\
 u(t)&r(t)
 \ea\,
 \right )\,
 \label{ondvabe}
 \ee
(with, say, positive $r(t)>0$) has the two eigenvalues
 \be
 g_\pm^{(2)}(t)= \pm
\sqrt{r^2(t)-u(t)v(t)}
 \label{formul}
 \ee
for which it is easy to find the boundary between the obsevable and
non-observable regimes. After a re-parametrization
 \ben
 u(t)=\frac{1}{2}\varrho(t)\,e^{\mu(t)}\,,
 \ \ \ \ \
 v(t)=\frac{1}{2}\varrho(t)\,e^{-\mu(t)}\,
 \een
we may recall Eq.~(\ref{formul}) and conclude that irrespectively of
the variation of the ``inessential" exponent $\mu(t)$ the system
will behave as unobservable at $\varrho(t)<-2 r(t)$, observable at
$-2 r(t) \leq \varrho(t) \leq 2 r(t)$ and unobservable again at
$\varrho(t)>2 r(t)$.

In the physical interval of $t \in (t_{initial},t_{final})$, i.e.,
for $\varrho(t)/[2\,r(t)] \in (-1,1)$, i.e., during all the
existence of our $N=2$ toy quantum Universe, the probabilistic
interpretation of its admissible states $|\psi^{(2)}\kt\in {\cal
H}^{(true)}(t)$ will be fully determined by the metric
$\Theta^{(2)}(t)>0$. Conditions (\ref{ultimo}) of the compatibility
of this metric with the geometry specified by the input operator
$\hat{G}(t)$ degenerate to the single constraint at $N=2$,
 \ben
    2 b(t) r(t) + u(t) a(t) + 2 b(t) u (t)+ v(t) a(t) + 2 b(t) v (t)=
    0\,.
 \een
In its light, up to an irrelevant overall factor the resulting
metric of the model becomes unique and solely defined in terms of
the (variable) matrix elements of $\hat{G}^{(2)}(t)$,
 \be
 \Theta^{(2)}(t)=
 \left (
 \begin{array}{cc}
 2r(t)&u(t)+v(t)\\
 u(t)+v(t)&2r(t)
 \ea\,
 \right )\,.
 \label{onidva}
 \ee
Eigenvalues $2r(t)\pm(u(t)+v(t))$ of this matrix must be both
positive so that we must keep $-2r(t)< u(t)+v(t)<2r(t)$ for
$t-t_{initial}$ small and positive as well as for $t-t_{final}$
small and negative.

We may fix another redundant degree of freedom by putting
$r(t)=1/2$. Then the  third parameter $\varrho(t)$ acquires the role
of a ``new time", with the simplest, linear exemplification
$\varrho_{lin}(t)=t/\alpha$. This reduction will lead to the BBC
identifications $t_{initial}=-\alpha$, $t_{final}=+\alpha$. In the
resulting model any spatial measurement before Big Bang as well as
after Big Crunch will only admit the  purely imaginary results
$g_\pm^{(2)}(t)$. Parameter $\mu(t)$ remains freely variable
controlling the probabilistic interpretation of the Universe in the
following three distinct dynamical regimes (cf. also
Fig.~\ref{fiedee}):

\begin{figure}[h]                     
\begin{center}                         
\epsfig{file=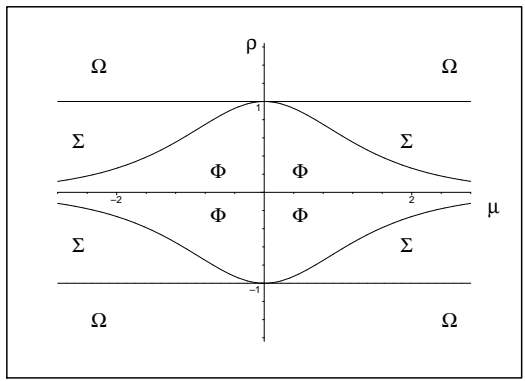,angle=270,width=0.5\textwidth}
\end{center}                         
\vspace{-2mm}\caption{Physical domain (marked by $\Phi$) in
$(\rho,\mu)-$plane.
 \label{fiedee}}
\end{figure}

 \begin{enumerate}

 \item
 for the times
  before Big Bang and after Big Crunch, i.e., in the unphysical domain
  with $ |\varrho| > 1$ (marked by symbol $\Omega$ in
  Fig.~\ref{fiedee})
 the spatial-point eigenvalues
  (\ref{formul}) stay  purely imaginary;
 the whole toy-model Universe remains
   unobservable;

 \item
 in the intermediate domain with
 $ 1 >|\varrho| >1/\cosh \mu $  (marked by symbol $\Sigma$ in
  Fig.~\ref{fiedee})
 the spatial-point eigenvalues get real but the toy-model Universe still
  cannot be given the probabilistic interpretation.
 The only candidate
 (\ref{onidva}) for the metric
 in ${\cal H}^{(true)}$
 remains indefinite or, in other words,
 no positive-definite metric
 becomes {\em simultaneously} compatible with input
 Hamiltonian (\ref{kinetie}) {\em and} with  input quantized geometry
 (\ref{formul});

 \item
 in the remaining and fully physical domain with
 $ |\varrho| <1/\cosh \mu $  (marked by symbol $\Phi$ in
  Fig.~\ref{fiedee}),
  {\em both}
 the given Hamiltonian  $H$ {\em and} the given  geometry
 $\hat{G}(t)$ become self-adjoint
 in ${\cal H}^{(true)}(t)$;
 the spatial-point eigenvalues stay real (= observable)
 while formula (\ref{onidva}) defines the unique, positive definite and
 adiabatically time-dependent metric.

 \end{enumerate}

 \noindent
On this background one has to impose the last, BBC-degeneracy
condition (\ref{acoll}) at $t=t_c$  reconfirming the expectations
that there are no free parameters at $N=2$ in general. Indeed,
Fig.~\ref{fiedee} shows that and why the BBC phenomenon may only be
consistently quantized at $\mu(t_c)=0$, i.e., just for the input
geometry $\hat{G}^{(2)}(t)$ characterized by the vanishing
asymmetry-parameter at $t=t_c$.

This is our first physics-mimicking observation which may also be
perceived as an encouragement of systematic study of the $N> 2$
models containing some variable parameters even at $t=t_c$. A
parallel, purely mathematical encouragement may be found in
Ref.~\cite{maxchain}. There, in different context, a very specific
generalization of our $\mu(t)=0$ model (denoted by symbol
$H^{(2)}_{(-1)}$ in {\em loc. cit.}) has been found tractable by
non-numerical means at all dimensions. Unfortunately, the number of
free parameters in these models proves too low for our present
purposes.

\subsection{Grid dimension $N=3$}

Hamiltonian (\ref{kinetie}) with $N=3$  possesses the three positive
time-independent eigenenergies $\varepsilon_0=2-2^{1/2}$,
$\varepsilon_1=2$ and $\varepsilon_2=2+2^{1/2}$. In combination with
its square
 \be
 H^2=
\left[ \begin {array}{ccc}
5&-4&1\\\noalign{\medskip}-4&6&-4\\\noalign{\medskip}1&-4&5\end
{array} \right]
 \ee
the insertion converts Eq.~(\ref{anza}) into the three-parametric
ansatz for the metric,
 \be
  \Theta^{(3)}=\left[ \begin {array}{ccc} a+2\,b+5\,c&-b-4\,c
  &c\\\noalign{\medskip}-b-4\,c&a+2\,b+6\,c&-b-4\,c\\
  \noalign{\medskip}c&-b-4\,c&a+2\,b+5\,c
\end {array} \right]\,.
 \label{metric3}
 \ee
All of the eigenvalues of the latter matrix, viz., the three
quantities
 $
 \theta_-=a+2\,b+6\,c-2^{1/2}\,(b+4\,c)$,
 $
 \theta_0= a+2\,b+4\,c$ and
 $
 \theta_+=a+
 2\,b+6\,c+2^{1/2}\,(b+4\,c)
 $
must be positive. This requirement specifies the boundary of the
domain of parameters ${\cal D}^{(physical)}$ in which the real and
symmetric matrix $\Theta^{(3)}$ may be treated as one of admissible
metrics in ${\cal H}^{(true)}$.

The reparametrization of $a=-2\,b-4\,c+\sqrt{2}\,\omega$ with
$\omega=\omega(a)>0$ reduces the definition of ${\cal
D}^{(physical)}$ to the elementary inequality $b < 2\sqrt{2}\omega$
and constraint
 \ben
  -\frac{\omega+b}{4+\sqrt{2}}<c<\frac{\omega-b}{4-\sqrt{2}}\,.
  \een
Inside these intervals we have to select parameters which make the
metric compatible with the operator $\hat{G}^{(3)}$. The not
entirely general, four-parametric classical-input-simulating choice
of the latter operator, viz.,
 \be
 \hat{G}^{(3)}=
 \left[ \begin {array}{ccc} -r&-u&-v\\\noalign{\medskip}u&0&-w
 \\\noalign{\medskip}v&w&r\end {array} \right]
 \label{operous}
 \ee
leads to the solvable secular equation for the observable grid
points $g$,
 \be
 -{g}^{3}+ \left( -{v}^{2}-{w}^{2}+{r}^{2}-{u}^{2} \right) {g}+{u}^{2}r-r{w}^{2}
 =0.
 \label{secul}
  \ee
In parallel, at $N=3$ the condition of ``hidden" Hermiticity of
operator (\ref{operous})  (i.e., Eq.~(\ref{ultimo})) degenerates to
the triplet of relations between the matrix elements of $
\Theta^{(3)}$ and $\hat{G}^{(3)}$,
 \be
 rb+4\,cr+2\,ua+4\,ub+11\,cu-vb-4\,cv-cw
 =0
 \ee
 \be
 -2\,cr-ub-4\,cu+2\,va+4\,vb+10\,cv-wb-4\,cw=0
 \ee
 \be
 -cu+2\,wa+4\,wb+11\,cw-vb-4\,cv+rb+4\,cr=0\,.
 \ee
The last line ceases to be linearly independent at $w=u$. The
reduced problem becomes easily solved in closed form,
 \be
 a={\frac {c \left( {r}^{2}+4\,ur+6\,{u}^{2}+2\,vr-3\,{v}^{2} \right) }{2
\,{u}^{2}+vr-{v}^{2}}}\,, \ \ \ \ w=u\,,
 \ee
 \be
 b=-2\,{\frac {c \left( ur+4\,{u}^{2}+2\,vr-2\,{v}^{2} \right) }{2\,{u}^{
2}+vr-{v}^{2}}}\,, \ \ \ \ w=u\,.
 \ee
Another simplification of the solution with $c=1$ and with $v=0$,
i.e., with the  tridiagonal input matrix $\hat{G}^{(3)}$ reads
 \be
 a={\frac { 6\,{u}^{2}+{r}^{2}+4\,ru  }{2{u}^{2}}}\,, \ \ \ \ w=u
 \,, \ \ \ \ v=0\,,\ \ \ \ c=1\,,
 \ee
 \be
 b=-{\frac { r+4\,u  }{u}}\,, \ \ \ \ w=u
 \,, \ \ \ \ v=0\,,\ \ \ \ c=1\,.
 \ee
After the latter reduction the triplet of the grid-point roots of
secular Eq.~(\ref{secul}) becomes particularly transparent,
 \be
 g_0=0\,,\ \ \ \ g_\pm = \pm \sqrt{r^2-2\,u^2}\,.
  \ee
For the time-independent particular choice of $r=\sqrt{2}$ the BBC
spatial singularity at $g_c=0$ is reached in the limit of $u\to
u_c=\pm 1$. For this reason we may treat $u$ as the updated,
rescaled time-variable at $N=3$.

The climax of the story is that the completion of the construction
of the probabilistic model, i.e., the search for a non-empty domain
${\cal D}^{(practical)}$ of positivity of the metric remains
non-numerical. The appropriate insertions imply that all of the
eigenvalues of the metric candidate $\Theta^{(3)}$ remain positive
for the one-parametric subfamily with fixed $v=0$, fixed
$r=\sqrt{2}$ and with the variable ``time" $w=u$ constrained to one
of the following two half-infinite intervals,
 \be
 u<-\frac{1}{1+\omega/\sqrt{2}}\,,\ \ \ \ \ \
  u > \frac{1}{1+\omega/\sqrt{2}}\,.
  \ee
As long as we have $\omega > 0$, the first one of these intervals
safely contains the instant $u_{initial}=- 1$ of Big Bang while the
second interval contains the Big-Crunch time $u_{final}=+1$.

In comparison with the preceding $N=2$ model, its updated $N=3$
descendant preserves the schematic pattern of the parametrization of
the operator of geometry as well as of its combination with
Hamiltonian $H^{(3)}$. A new qualitative feature emerges since at
$N=2$ the two input observables already determined the admissible
metric completely. At $N=3$ one of the parameters [viz.,
$\omega=\omega(t)$] remains variable and may be adjusted to some
additional phenomenological requirements (the deeper discussion of
this problem of ambiguity of $\Theta$ as presented in
Ref.~\cite{Geyer} should be consulted in this context).

\section{Evolution in time near $t=t_c$\label{thebeat}}

The study of our family of toy models at higher numbers of grid
points $N$ would require the use of the standard numerical and
computer-assisted tools of linear algebra. The quick growth of the
number $N^2$ of available free parameters in the input geometry
matrix $\hat{G}$ would make such a study unnecessarily extensive.
Thus, a concrete phenomenological motivation narrowing the choice of
the input matrices $\hat{G}^{(N)}(t)$ would be welcome.

In our present methodical considerations we may only try to separate
the set of the matrix elements $\gamma_{jk}$ into its ``important"
and ``less essential" subsets. One of the methods of such a
reduction of the input information is provided by the possibility of
the elementary-rotation reduction of a general finite-dimensional
matrix to its ``canonical" Hessenberg form \cite{Wilkinson}. In this
sense let us now admit just the special, tridiagonal form of
matrices $\hat{G}^{(N)}$ containing $3N-2$ ``most important" real
parameters.

We expect that due to the tridiagonal structure of our toy
Hamiltonian (\ref{kinetie}) the number of  independent items in the
metric-compatibility condition (\ref{ultimo}) will be much lower
than predicted by our original upper estimate $N(N-1)/2$ based on
the mere antisymmetry of the general matrix expression. In such a
reduced setting the $N-$plet of constraints (\ref{acoll}) of the
necessary spectral reality as well as the BBC degeneracy condition
(\ref{collide}) will play a much more decisive role, indeed.
Nevertheless, we believe that the use of the tridiagonal matrices
$\hat{G}^{(N)}$ will still leave some of their parameters
unrestricted so that, from the point of view of physics, no unstable
fine-tuning will be required even after such a drastic
simplification of the underlying mathematics.

Our final sample of solvable examples may clarify this point.

\subsection{$N=4$ model and BBC degeneracy  at $t=t_c$}

The use of the simplest four-parametric toy model with $s>r>0$ in
 \be
 \hat{G}^{(4)}=
 \left[ \begin {array}{cccc} -s&-u&0&0\\
 \noalign{\medskip}u&-r&-p&0\\
 \noalign{\medskip}0&p&r&-u\\
 \noalign{\medskip}0&0&u&s
\end {array} \right]
 \label{geo4}
 \ee
preserves the exact solvability of the secular equation,
 \be
{g}^{4}+ \left( -{s}^{2}+2\,{u}^{2}+{p}^{2}-{r}^{2} \right)
{g}^{2}+{r}^{2}{s}^{2}+2\,sr{u}^{2}-{p}^{2}{s}^{2}+{u}^{4}=0\,.
\label{colied}
 \ee
For $t \in (t_{initial},t_{final})$ all of its roots given by the
standard elementary formulae must be real. Thus, not only that the
first coefficient in Eq.~(\ref{colied}) must be non-positive, i.e.,
 \be
 {r}^{2}+{s}^{2}\geq 2\,{u}^{2}+{p}^{2}
 \ee
but also we must demand that
 \be
 \left ({r}{s}+{u}^{2}\right )^2 \geq {p}^{2}{s}^{2}\,.
 \ee
The third requirement must guarantee the non-negativity of  the
discriminant of our quadratic equation,
 \be
 {s}^{4}-4\,{s}^{2}{u}^{2}+2\,{p}^{2}{s}^{2}
 -2\,{r}^{2}{s}^{2}+4\,{p}^{2}{u}^{2}-4\,{r}^{2}{u}^{2}+{p}^{4}
 -2\,{p}^{2}{r}^{2}+{r}^{4}
 -8\,sr{u}^{2}\geq 0\,.
 \ee
This relation may be further simplified as follows,
 \be
 (p^2+s^2-r^2)^2\geq 4\, u^2\,[(s+r)^2-p^2]\,.
 \label{thertsec}
 \ee
The BBC phenomenon will be characterized by the quadruple confluence
of the real roots $g_k$ which is only possible when $g_c=0$. Then,
constraint (\ref{thertsec}) becomes redundant and we get two
conditions at $t=t_c$, viz.,
 \be
 {s}^{2}_c+{r}^{2}_c=2\,{u}^{2}_c+{p}^{2}_c
 \label{thefir}
 \ee
and
 \be
 (r_cs_c+u^2_c)^2=p^2_c\,s^2_c\,.
 \label{thesec}
 \ee
The elimination of $p^2_c$ defined by the former relation
(\ref{thefir}) leaves us with the three real BBC parameters
constrained by the single equation
 \be
 2r_cs_cu^2_c+u^4_c+2s^2_cu^2_c-s^4_c=0\,.
 \ee
Most easily we may keep $s_c$ and $u_c$ as two freely variable
parameters and eliminate
 \be
 r_c = r_c(s_c,u_c)=-s_c -\frac{s_c}{2}
 \left [
\frac{u^2_c}{s^2_c}- \frac{s^2_c}{u^2_c}
 \right ]\,.
 \ee
This means that using Eq.~(\ref{thesec}) we have to define
 \be
 p_c=p_{\pm c}(s_c,u_c) =\pm [r_c(s_c,u_c)+u^2_c/s_c]\,.
 \ee
In place of independent variable $u_c$ an alternative real parameter
$\varrho$ may be used in a  reparametrization
 \be
 u_c=u_{\pm c}(s_c,\varrho) =\pm s_c\,e^{-\varrho}\,.
 \ee
This finally simplifies the form of the quantity
 \be
 r_c = r_c(s,\varrho)={s}_c\,[-1+{\rm sinh}\,2\,\varrho]
 \,.
 \ee
We may conclude that the existence of the BBC phenomenon in our
$N=4$ model with $s>r>0$ will be guaranteed whenever the variability
of $\varrho$ is restricted to the interval where $ {\rm
sinh}\,2\,\varrho \in (1, 2)$.

\subsection{Evolution  near $t=t_c$ }

Let us now return to Ref.~\cite{chain} where we analyzed the
properties of a four-dimensional matrix which coincides with our
geometry operator (\ref{geo4}) at the constant sample values of
$r(t)=1$ and $s(t)=3$ corresponding to the special and,
incidentally, BBC-compatible value of $ {\rm sinh}\,2\,\varrho =
4/3$. In different context, a very specific time-dependence of the
remaining two variable matrix elements has been postulated there,
 \be
 u(t)=-\sqrt{3-3t-3Bt^2}\,,\ \ \ \
 p(t)=-\sqrt{4-4t-4A t^2}\,.
 \label{paira}
 \ee
This form of time-dependence of the system serves our present
purposes well. Once we choose $A=B=-1/2$ we obtain the standard
global BBC scenario in which the observable Universe exists strictly
inside the whole interval of times $t\in (t_{initial},t_{final})$
with $t_{initial}=0$ and $t_{final}=2$. As we already explained,
however, without a deeper insight into the (presumably, covariantly
described) dynamics of similar systems, the explicit time-dependence
of the observable quadruplet of grid points as given by
Eq.~(\ref{paira}) only keeps its good physical meaning in some very
short intervals of the ``classical" continuous times near
$t_{initial}$ or $t_{final}$.

%
%
%

The same comment applies also in the case of the alternative choice
of $A=B=+1/2$ which leads to the permanently expanding Universe.
Within the framework of our toy model the size of this ``Universe"
is just an asymptotically linear function of time.

%

We may conclude that the available menu of qualitative physical
predictions remains sufficiently sensitive to the variations of our
dynamical ``input" assumptions. The bad news is that the choice of
the mere two parameters (\ref{paira}) in the input
$\hat{G}^{(4)}(t)$ where $r(t)=1$ and $s(t)=3$ (i.e., the lack of
necessary parameters rendering Eq.~(\ref{ultimo}) valid) already
makes the resulting metric $\Theta^{(4)}(t)$ either incompatible
with the Hamiltonian $H^{(4)}$ of Eq.~(\ref{kinetie}) or,
alternatively, compatible with this Hamiltonian at a single, {\em
BBC-incompatible} time $t =t_{fixed}\neq t_c$.

This is a not too essential cloud which has its silver linen since
the corresponding lengthy calculations (which we omit here) reveal
that the distance between $t_{fixed}$ and $t_c$ proves unexpectedly
small (in fact, of the order of $10^{-2}$ in our units). This
indicates that any amended (i.e., necessarily, more-parametric)
BBC-compatible input matrix $\hat{G}^{(4)}(t)$ will be not too
different from its imperfect but still sufficiently transparent
present  illustrative $r_c=1$ solvable example where we followed
Ref.~\cite{chain} and choose $ {\rm sinh}\,2\,\varrho = 4/3$.

\section{Comments and summary \label{therest} }

In our paper we detected a gap in the argumentation denying the
compatibility of the BBC phenomena with quantum mechanics. Our main
assertion was that after an appropriate amendment of the
representation of states the Big Bang/Crunch (BBC) phenomenon may
remain fully compatible with the very standard textbook quantum
theory known from traditional textbooks \cite{Messiah}.

In our non-covariant, purely quantum-mechanical toy-model simulation
of an exploding or collapsing Universe an entirely elementary
Hamiltonian $H=H^\dagger$ was complemented by a less trivial though
still highly schematic cryptohermitian observable $\hat{G} \neq
\hat{G}^\dagger$ which was required to represent a time-dependent
spatial geometry near a hypothetical Big Bang/Big Crunch
singularity.

An {\em alternative} operation of Hermitian conjugation has been
introduced serving as an {\em ad hoc} definition of an amended,
physical Hilbert space of states ${\cal H}^{(true)}$. This enabled
us to keep {\em both} the observables $H$ and $\hat{G}$ self-adjoint
{\em strictly inside} a finite interval of time $t \in
(t_{initial},t_{final})$. Beyond its boundaries (i.e., before Big
Bang or after Big Crunch) the eigenvalues of $\hat{G}$ were allowed
to get complex so that the Universe ceased to be observable.

We argued that the presented form of a purely quantum-mechanical
collapse of our toy-model Universe at $t= t_{initial/final}$ was in
fact mediated by the introduction of the ``true" or
``self-consistent" {\em manifestly time-dependent}
Hermitian-conjugation operation. In such a setting particular
attention has been paid to the ambiguity of the choice of the inner
product as discussed in Refs.~\cite{Geyer,timedep}.

Many questions have been skipped as inessential for our present,
predominantly methodical purposes. Naturally, these questions will
re-emerge immediately in any phenomenologically oriented
considerations in which

\begin{itemize}

\item
(a) a more specific form of the adiabatically time-dependent input
matrix elements $\gamma_{jk}(t)$ of the operator $\hat{G}$ would be
deduced from the classical general relativity theory, say, on the
basis of some suitable version of the principle of correspondence;

\item
(b) a number of other observables (say, $\hat{F}_1$, $\hat{F}_2$,
$\ldots$) would be introduced as reflecting, say, the presence of
some matter fields;

\item
(c) the dimension $N$ which characterizes the discretization
approximation would be sent to its infinite, continuous-space limit;

\item
(d) a realistic, three-dimensional measurable space would be
considered;

\item
(e) at least an approximate version of the Lorentz
special-relativistic covariance of kinematics would be taken into
account, etc.

\end{itemize}

 \noindent
Our present discrete odd$-N$ model might be also interpreted as
allowing the existence of a zero-dimensional observable Universe
before Big Bang and/or after Big Crunch. Thus, in a more realistic
three-dimensional Universe one could proceed in the highly
speculative spirit of Refs.~\cite{Penrose,PenroseII} and conjecture
that our Universe might just change its dimension during Big Bang
and/or Big Crunch.

Up to similar exceptions we tried here to avoid all of the
speculative considerations. Instead, we presented just a few purely
formal arguments based on the analysis of a few elementary models.
Our results may be briefly characterized as a demonstration of
tractability of the quantization of systems which seem to exhibit a
``catastrophic", BBC-resembling time-evolution behavior in their
classical models. The main sources of our proposed systematic
approach to quantization of such systems may most briefly be
summarized as lying in the following four assumptions of

\begin{itemize}

\item
(i) the availability of some external, non-quantum information about
the system exemplified here by the expected knowledge of the
``input" matrices $\hat{G}=\hat{G}(t)$) plus  $H\neq H(t)$ and also,
perhaps, $\hat{F}_1(t)$, $\hat{F}_2(t)$, $\ldots$;

\item
(ii) the availability of some theoretical background for decisions,
say, between the admissibility \cite{fund} and inadmissibility
\cite{cubic} of a fundamental length in the model;

\item
(iii) the feasibility of calculations; as long as we decided to
admit nontrivial metrics $\Theta(t)\neq I$, this apparently purely
formal requirement proves of paramount importance as limiting, e.g.,
the range of practical applicability of perturbation expansions
\cite{Jones} or of the Moyal-bracket recipes \cite{Scholtz} etc;

\item
(iv) the feasibility of making the metric $\Theta$ compatible with
{\em two and more} cryptohermitian observables; up to now there
existed not too many constructions of this type \cite{ccSIGMAbe};
even in our present paper we considered just $H \neq H(t)$.
Moreover, we did not dare to move beyond the mere adiabatic
dynamical regime.

\end{itemize}


\subsection*{Acknowledgments}

Work supported by the GA\v{C}R grant Nr. P203/11/1433, by the
M\v{S}MT ``Doppler Institute" project Nr. LC06002 and by the
Institutional Research Plan AV0Z10480505.

\newpage

\section*{Appendix A. BBC degeneracy at  $N=5$}

The four-parametric ansatz with $s \geq  r \geq 0$,
 \be
 \hat{G}^{(5)}=
  \left[ \begin {array}{ccccc} -s&-u&0&0&0\\\noalign{\medskip}u&-r&-p&0
&0\\\noalign{\medskip}0&p&0&-p&0\\\noalign{\medskip}0&0&p&r&-u
\\\noalign{\medskip}0&0&0&u&s\end {array} \right]
 \ee
leads to the secular equation
 \be
-{g}^{5}+ \left( {s}^{2}-2\,{u}^{2}-2\,{p}^{2}+{r}^{2} \right)
{g}^{3}+ \left( -{r}^{2}{s}^{2}-2\,{u}^{2}rs+2\,{p}^{2}{s}^{2}-
2\,{p}^{2}{u}^{2}-{u}^{4} \right) g =0
 \ee
with one root $g_0=0$ and  four roots $g_{\pm,\pm}$ given by the
standard formulae.  
%
%
%
%
%
%

At $t=t_c$ we may apply the same sequence of manipulations as used
at $N=4$  leading to the modified pair of BBC-degeneracy constraints
 \be
 {s}^{2}_c+{r}^{2}_c=2\,{u}^{2}_c+2\,{p}^{2}_c
 \label{thefir5}
 \ee
 and
 \be
 (r_cs_c+u^2_c)^2=2\,p^2_c\,\left (s^2_c - u^2_c\right )\,.
 \label{thesec5}
 \ee
We have to keep $u^2_c\leq s^2_c$, i.e.,
 \be
 u_c=u_{\pm c} (s_c,\varrho) = \pm  s_c\,e^{-\varrho}\,,\ \ \ \ \ \ \ \varrho
 \geq 0\,.
 \ee
The elimination of the ``redundant" $p^2_c$ defined by relation
(\ref{thefir5}) leads to the single constraint
 \be
 u^2_c\,(r_c+s_c)^2=\left (s^2_c - u^2_c\right )^2\,
 \ee
which provides the two alternative definitions of the second
dependent parameter
 \be
 r_c=r_{\pm c}=-s_c\pm s_c\,\left (\frac{s_c}{u_c}-\frac{u_c}{s_c}\right )\,.
 \ee
As long as we decided to require that $s_c \geq r_c\geq 0$, we
arrive at the unique prescription
 \be
 r_c = r_c (s_c,\varrho)={s}_c\,[-1+ {\rm sinh}\, \varrho]\,,
 \ \ \ \ \  1 \leq {\rm sinh}\,\varrho \leq 2
 \,.
 \ee
Once more we may  recall Ref.~\cite{chain} and find there the
special case of our five-dimensional geometry-operator matrix
(\ref{geo4}) with $r(t)=2$ and $s(t)=4$, i.e., with $ {\rm
sinh}\,\varrho = 3/2$. Borrowing again the specific time-dependence
of the matrix elements from the same reference,
 \be
 u(t)=-\sqrt{4-4t-4 B t^2}\,,\ \ \ \
 p(t)=-\sqrt{6-6t-6 A t^2}\,, \ \ \ \ \ t_c=0\,,
 \label{pairbe}
 \ee
we obtain the time-dependent $N=5$ spectrum $\{g_k(t)\}$ resembling
the $N=4$ pattern.

%

The news are that in a way typical for the odd dimensions $N$ one of
the roots (viz., the time-independent $g_0(t)=0$) remains real at
all times. Once we choose $A=B=-1/2$ we obtain the global BBC
scenario in which the observable Universe exists for all times but
it is one-dimensional for $t\in (t_{initial},t_{final})$ and
zero-dimensional for $t\notin [t_{initial},t_{final}]$ (we have
$t_{initial}=0$ and $t_{final}=2$ in our units).

Within our present restricted perspective provided by the mere
standard quantum mechanics the complexification of the eigenvalues
is precisely what is meant by the words ``non-observable". A good
textbook illustration is provided by the Coulomb field in the Dirac
equation in the superstrong-coupling regime where the sudden
complexification of the energies mimics the moment of the sudden
emergence of the many-body physics via the suddenly opened channel
admitting the creation of particle-antiparticle pairs \cite{Bohr}. A
thinking by analogy could equally well apply in the present model
where one measures the coordinates and where their complexification
might be also interpreted as mimicking a decrease of the dimension
of the space in a continuous limit $N \to \infty$. Of course, the
decrease of the dimension is not the only possibility. One could
even develop many other {\em ad hoc} toy models supporting
alternative scenarios supporting, say, a complete survival of the
observability (without any change of dimension - this would be
typical, say, for the cyclic cosmologies of Ref.~\cite{CCC}) or even
an increase of the dimension (which could even lead, in cosmology,
to certain truly speculative darwinistic-sounding concepts
\cite{Stolin}).

\end{document}